\documentclass[prl,twocolumn,amsmath,showpacs]{revtex4}
\usepackage{graphicx}
\usepackage{amssymb}
\usepackage{citesort}
\usepackage{stmaryrd}

\begin{document}

\title{Essential optical states in $\pi$-conjugated polymer thin films}

\author{Zhendong Wang}
\affiliation{Department of Physics, University of Arizona
Tucson, AZ 85721}
\author{Alok Shukla}
\affiliation{Physics Department, Indian Institute of Technology, Powai, Mumbai 400076, India}
\author{Sumit Mazumdar}
\affiliation{Department of Physics, University of Arizona
Tucson, AZ 85721}
\date{\today}
\begin{abstract}
We develop a theory of the electronic structure and photophysics of interacting chains
of $\pi$-conjugated polymers to understand the differences between solutions and
films. 
While photoexcitation generates only the exciton in solutions, the optical 
exciton as well as weakly allowed
excimers are generated in films.
Photoinduced absorption in films is primarily from the lowest excimer. We
are also able to explain peculiarities associated with photoluminescence,
including delayed photoluminescence and its quenching by electric field.
\end{abstract}
\pacs{42.70.Jk, 71.35.-y, 78.20.Bh, 78.30.Jw}
\maketitle

The photophysics of dilute solutions and thin films of $\pi$-conjugated polymers (PCPs)
are often remarkably 
different \cite{Rothberg06,Arkhipov04,Schwartz03}. Solutions exhibit
behavior characteristic of single strands, and the different behavior of films
is due to interchain interaction and disorder. Microscopic understanding
of the effects of interchain interaction has remained incomplete
even after intensive investigations. 
As discussed below, the experimental results
themselves are controversial.
Further, while theoretical
works on interchain interactions have examined a variety of issues  \cite{theory1}, 
few authors \cite{theory2} have investigated the interchain species that 
dominate thin film photophysics \cite{Rothberg06,Arkhipov04,Schwartz03}. 

The key controversy
in PCP films involves the apparent generation of polaron-pairs
(bound electron-hole pairs on neighboring chains)
or even free polarons
upon photoexcitation. The reduced 
photoluminescence (PL) of films,  
nonexponential time decay of PL, delayed PL
lasting until milliseconds, and electric-field induced 
quenching of the same 
are cited as evidence for the polaron-pair \cite{Rothberg06,Arkhipov04,Schwartz03}. 
The mechanism of their formation \cite{Basko02} is controversial.
Transient absorptions in films
are also not understood. Two distinct ultrafast photoinduced absorptions (PAs)  
are seen in solutions as well as in films with weak
interchain interactions, such as dioctyloxy-poly-paraphenylenevinylene 
(DOO-PPV) \cite{Frolov00}.
The low energy PA$_1$ appears at a threshold energy of 0.7 eV and has a peak at
$\sim$ 1 eV, while the higher energy PA$_2$ occurs at $\sim$ 1.3--1.4 eV. Comparison
of PA and PL decays \cite{Frolov00} and other nonlinear spectroscopic measurements 
\cite{Liess97} have
confirmed that these PAs 
are from the 1B$_u$
optical exciton, in agreement with theoretical work on 
PCP single chains \cite{theory-nlo,Shukla03}. In contrast, PA and PL
decays in PCPs with significant
interchain interactions (for e.g., 
poly[2-methoxy,5-(2'-ethyl-hexyloxy)-p-phenylenevinylene], MEH-PPV) are 
uncorrelated \cite{Rothberg06,Yan94}. PAs in these are assigned
to the polaron-pair \cite{Rothberg06,Yan94}. 

Recent experiments have contributed further to the mystery.
Sheng {\it et al.} have extended femtosceond (fs) PA measurements
to previously inaccessible wavelengths \cite{Sheng07}.
The authors have detected a new relatively weak PA at $\sim$ 0.35--0.4 eV in PCP 
films \cite{Sheng07}, which they label P$_1$.
The authors ascribe P$_1$ to free polaron absorption, and suggest 
instantaneous photogeneration of both polarons and excitons.
Such branching of photoexcitations would be in agreement
with previous claim of the observation of infrared active vibrations in MEH-PPV
in fs time \cite{Miranda04}, but is difficult to reconcile with
(i) the large exciton binding energies deduced from PA$_1$ energy 
\cite{Frolov00,Liess97,theory-nlo}, and (ii) microwave conductivity measurements 
\cite{Dicker04} and THz spectrscopy \cite{Hendry05},
which find negligible 
polaron generation in solutions and films.
Interestingly, samples that exhibit P$_1$ absorption also exhibit induced
absorptions PA$_1$ and PA$_2$ \cite{Sheng07}, in apparent contradiction to the polaron-pair
picture \cite{Rothberg06,Yan94}
(unless PAs from the latter and the exciton occurred at the same energies).
Taken together, the above experiments
indicate the strong need for systematic theoretical work on 
interacting PCP chains.

In the present Letter, starting from a microscopic $\pi$-electron Hamiltonian,
we determine the complete energy spectrum of interacting
PCP chains. We find a 1:1 correspondence between 
the ``essential states'' that determine the photophysics of 
single strands \cite{theory-nlo,Shukla03} and the dominant excited states,
including {\it excited interchain species}, in films.
We are able to explain consistently, (i) the branching of 
photoexcitations, 
(ii) the peculiarities associated with PL, 
and (iii) ultrafast PA over the complete experimental wavelength region for PCP films.

Our calculations are within an extended two-chain Pariser-Parr-Pople Hamiltonian \cite{PPP}
$H = H_{intra}+H_{inter}$, where $H_{intra}$ and $H_{inter}$ correspond to intra-
and interchain components, respectively. $H_{intra}$ is written as,
\begin{multline}
\label{H_PPP}
H_{intra} = - \sum_{\nu, \langle ij \rangle, \sigma} t_{ij}
(c_{\nu,i,\sigma}^\dagger c_{\nu,j,\sigma}+ H.C.) \\+
U \sum_{\nu,i} n_{\nu,i,\uparrow} n_{\nu,i,\downarrow} 
+ \sum_{\nu,i<j} V_{ij} (n_{\nu,i}-1)(n_{\nu,j}-1)
\end{multline}
where $c^{\dagger}_{\nu,i,\sigma}$ creates a $\pi$-electron of spin $\sigma$ on
carbon atom $i$ of oligomer $\nu$ ($\nu$ = 1, 2), $n_{\nu,i,\sigma} = 
c^{\dagger}_{\nu,i,\sigma}c_{\nu,i,\sigma}$ is
the number of electrons on atom $i$ of oligomer $\nu$ with spin $\sigma$ and
$n_{\nu,i} = \sum_{\sigma}n_{\nu,i,\sigma}$ is the total number of electrons on atom
$i$. The hopping matrix element
$t_{ij}$
is restricted to nearest neighbors and
in principle can contain electron-phonon interactions, although
a rigid bond approximation is used here. $U$ and $V_{ij}$ are the
on-site and intrachain intersite Coulomb interactions.
We parametrize $V_{ij}$
as \cite{Chandross97}
\begin{equation}
\label{parameters}
V_{ij}=\frac{U}{\kappa\sqrt{1+0.6117 R_{ij}^2}} 
\end{equation}
where $R_{ij}$ is the distance between carbon atoms $i$ and $j$ in
\AA, and $\kappa$ is the dielectric
screening along the chain due to the medium. Based on previous work \cite{Chandross97}
we choose $U$ = 8 eV and $\kappa$ = 2.
We write $H_{inter}$ as
\begin{align}
\label{inter}
H_{inter} &= H_{inter}^{1e}+H_{inter}^{ee} \\
H_{inter}^{1e} &= -t_{\perp}\sum_{\nu <
\nu^{\prime},i,\sigma}(c^{\dagger}_{\nu,i, \sigma}
c_{\nu^{\prime},i,\sigma} + H.C.)  \\
H_{inter}^{ee}&=\frac{1}{2}\sum_{\nu < \nu^{\prime},i,j} V_{ij}^{\perp}(n_{\nu,i} -1)(n_{\nu^{\prime},j} - 1)
\end{align}

In the above, $t_{\perp}$ is restricted to
nearest interchain neighbors.
We choose $V_{ij}^{\perp}$ as in Eq. \ref{parameters}, with
a background dielectric constant $\kappa_{\perp} \leq \kappa$ \cite{theory2}.

To get a physical understanding
of the effect of $H_{inter}$, we begin with the case of two ethylene molecules
in the molecular orbital (MO) limit $U=V_{ij}=0$, placed
one on top of the other such that the overall structure has a center of inversion.
The ethylene MOs are written as,
\begin{equation}
\label{a_dagger}
a_{\nu,\lambda,\sigma}^\dagger = {1 \over \sqrt{2}}
[c_{\nu,1,\sigma}^\dagger + (-1)^{(\lambda - 1)}c_{\nu,2,\sigma}^\dagger]
\end{equation}
where $\lambda=1(2)$ corresponds to the bonding (antibonding) MO.
The spin singlet one-excitation space for the two molecules consists of the four
spin-bonded valence bond (VB) diagrams shown in Fig.~1(a). We refer to the
two intramolecular excitations
$|exc1 \rangle$ and $|exc2 \rangle$
as excitons (we ignore for the moment that 
true excitons require nonzero $U$ and $V_{ij}$); the other two,
$|P_1^+P_2^- \rangle$ and $|P_1^-P_2^+ \rangle$, consist of charged molecules
and are the polaron-pairs. The exciton and polaron-pair wavefunctions are given by,
\begin{align}
\label{VB}
|exc1 \rangle = {1 \over \sqrt{2}}a_{2,1,\uparrow}^\dagger a_{2,1,\downarrow}^\dagger (a_{1,1,\uparrow}^\dagger a_{1,2,\downarrow}^\dagger -
a_{1,1,\downarrow}^\dagger a_{1,2,\uparrow}^\dagger)|0 \rangle \\
|P_1^+P_2^- \rangle = {1 \over \sqrt{2}}a_{2,1,\uparrow}^\dagger a_{2,1,\downarrow}^\dagger (a_{1,1,\uparrow}^\dagger a_{2,2,\downarrow}^\dagger -
a_{1,1,\downarrow}^\dagger a_{2,2,\uparrow}^\dagger)|0 \rangle 
\end{align}

\begin{figure}
 \centering
 \includegraphics[clip,width=3.375in]{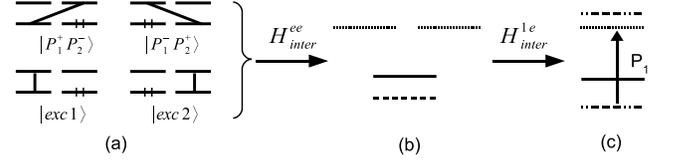}
 \caption{(a). The one-excitation space of two weakly
interacting oligomers. For each oligomer one bonding and one antibonding MO is shown,
with singlet bonds between singly occupied MOs. (b) and (c). Eigenstates of
$H_{inter}^{ee}$ and total $H_{inter}$, respectively. 
Solid lines, even parity exciton; dashed line, odd parity
exciton; dot-dashed lines, excimers; dotted lines, polaron-pairs.
The characters of the final four states are the
same for arbitrary PCPs, even for nonzero $U$ and $V_{ij}$ and
excited states far from the optical edge.
The P$_1$ photoinduced absorption \cite{Sheng07} from the lowest excimer is indicated.  
}
\end{figure}

Nonzero $H_{inter}$ mix these pure states
to give {\it excimers}.
Consider first the $H_{inter}^{1e}=0$ limit.
Matrix elements of $H_{inter}^{ee}$ are zero between $|P_1^+P_2^- \rangle$ and 
$|P_1^-P_2^+ \rangle$ but nonzero between $|exc1 \rangle$ and $|exc2 \rangle$, 
indicating that
while the polaron-pair states are degenerate for $H_{inter}^{ee} \neq 0$,
the exciton states form new nondegenerate states $|exc1 \rangle \pm |exc2 \rangle$
(see Fig.~1(b)).
The dipole operator $\pmb{\mu} = e\sum_{\nu,i}\pmb{r}_{\nu,i}(n_{\nu,i}-1) $, 
where $\pmb{r}_{\nu,i}$ gives the location of atom
$i$ on oligomer $\nu$,
couples the ground state to only the even parity 
exciton state.
We now switch on $H_{inter}^{1e}$,
which mixes $|exc1 \rangle-|exc2 \rangle$ and $|P_1^+P_2^- \rangle-|P_1^-P_2^+ \rangle$,
to give the two excimer states in Fig.~1(c). From the expression
for $\pmb{\mu}$,
both excimers are dipole-forbidden from the ground state. 
The optical exciton $|exc1 \rangle + |exc2 \rangle$
and the polaron-pair $|P_1^+P_2^- \rangle+|P_1^-P_2^+ \rangle$ are not affected
by $H_{inter}^{1e}$ in this symmetric case.
We make an important observation: the transverse component of $\pmb{\mu}$,
perpendicular to the molecular axes, couples the two
excimer states with the polaron-pair state, 
indicating {\it allowed optical transition
from the lower excimer to the polaron-pair.}

We now go beyond the two ethylenes and make the following observations.

(i) Fig.~1 applies also to the $U=V_{ij}=0$ limit of
identical oligomers of arbitrary PCPs facing one another.
The lowest  
one-excitations, across the highest occupied and lowest unoccupied MOs
of the oligomer pairs, are split by $H_{inter}$ to give the same four
eigenstates of Fig.~1. Wu and Conwell, and Meng had concluded the same
for a simpler $H^{ee}_{inter}$ \cite{theory2}.
We have confirmed this for $H_{inter}$ in Eqs. 3-5 from configuration
interaction (CI) calculations involving the configurations of Fig.~1 for long
PPV oligomers.
To identify wavefunctions as polaron-pair, excimer, etc., 
we choose an  orbital set consisting of the canonical 
Hartree-Fock orbitals of the individual molecular
units, and perform CI calculations using these localized MOs.
The localized basis allows calculations of ionicities of individual oligomers.
The expected ionicities are 0 and 1 for the exciton and the polaron-pair, respectively,
and fractional for the excimers. 

(ii) Fig.~1 applies also to {\it higher energy excitations} involving 
{\it arbitrary} pairs of MOs of PCP oligomers. Note that interchain
species here are in their {\it excited states}. 
Again, we have verified this from CI calculations involving highly 
excited pairs of MO configurations of PPV oligomers, using the localized basis.

\begin{table}
\caption{SCI excited states of two symmetrically
placed 8-unit PPV oligomers for $\kappa_{\perp}=2$,
$t_{\perp}=0.03$ and 0.1 eV 
(numbers in parentheses are for $t_{\perp}=0.1$ eV.)
Here $j$ and $E_j$ are quantum numbers (without considering symmetry)
and energy, respectively. Ionicity is the charge on the
chains. 
The states are arranged not according to their energies, but 
according to the manifolds they belong to (see text). The
$\mu_{G,j}$ and $\mu_{i,j}$ are the dipole couplings
between the ground state and state $j$, and between excited states,
respectively.}
\begin{ruledtabular}
\begin{tabular}{ccccc}
   $j$ & $E_j$ (eV) & Ionicity  & $\mu_{G,j}$ & $\mu_{i,j}$ \\ \hline
  2 (2)	   & 2.77 (2.67) & 0.06  (0.26) & 0    & \textemdash \\
  3 (4)    & 2.83 (2.81) & 0    (0)    & 6.52 (6.52) & \textemdash       \\
  5 (5)    & 3.00 (3.00) & 1  (1)  & 0 & 0.99(2.04)$^a$     \\
  6 (8)    & 3.02 (3.12) & 0.94 (0.74) &  0   & \textemdash       \\ 
  \\
  4 (3)	   & 2.92 (2.81) & 0.08 (0.29) & 0  & \textemdash \\
  7 (7)    & 3.06 (3.06) & 0   (0)    & 0  & \textemdash   \\
  8 (9)    & 3.12 (3.12) & 1 (1)  &  0 & \textemdash  \\
  9 (10)   & 3.14 (3.24) & 0.92 (0.64)& 0  & \textemdash      \\ 
  \\
  14 (11)   & 3.39 (3.26) & 0.18 (0.38) & 0  & 6.66 (6.91)$^b$\\
  16 (15)   & 3.42 (3.42) & 0    (0)    & 0  & 6.83 (6.83)$^{b,c}$\\
  18 (19)   & 3.52 (3.46) & 1  (1)  &  0 & 7.68 (7.68)$^b$\\
  20 (26)   & 3.55 (3.67) & 0.82 (0.55) & 0  & 7.41 (6.69)$^b$\\  
\end{tabular}
\end{ruledtabular}
\footnotetext{$^a$i=2. $^b$All dipole couplings are with states in lowest
manifold near 1B$_u$ with the same character (see text). $^c$ The mA$_g$.}
\end{table}

\begin{table}
\caption{SCI excited states near the optical exciton of 
two-chain PPV with 5 and 7-units for $\kappa_\perp=2$, 
$t_\perp=0.1$ eV.
}
\begin{ruledtabular}
\begin{tabular}{ccccc}
   j & $E_j$ (eV) & Ionicity & $\mu_{G,j}$ & $\mu_{2,j}$  \\ \hline
  2     & 2.74  & 0.23   & 1.45  & \textemdash \\
  3     & 2.89  & 0.07   & 4.99  & 0     \\
  4     & 3.04  & 0.14  & 2.07  & 0     \\
  5     & 3.07  & 1.00  & 0     & 2.03    \\ 
  6     & 3.16  & 0.45  & 0.31  & 0     \\
  7     & 3.21  & 0.53  & 0.57  & 0      \\
\end{tabular}
\end{ruledtabular}
\end{table}

(iii) The results of (i) and (ii) indicate that for $U=V_{ij}=0$ but
$H_{inter} \neq 0$, the two-chain energy spectrum consists of a series of 
overlapping energy manifolds, with each manifold containing an
exciton, a polaron-pair and two excimers, as in Fig.~1. 
We speculate
that {\it similar energy manifolds persist for nonzero $U, V_{ij}$.}
We have verified this conjecture within the single-CI
(SCI) approximation for the {\it complete two-chain Hamiltonian,} including {\it all}
one-excitations, and using the localized MO basis set.
We have summarized our SCI results in Table I for 
two interacting symmetrically placed parallel PPV oligomers at a distance of 0.4 nm.
Identification of excitons, polaron-pairs and excimers
from their ionicities is
possible even at higher energies. Table I also lists relevant transition 
dipole couplings. 
PA from the lowest
excimer ($i=2$) to the lowest polaron-pair ($j=5$) is still allowed. PA in single-chain PCPs is from the
1B$_u$ exciton to a
dominant two-photon exciton labeled the mA$_g$ \cite{theory-nlo}, and to an even higher
energy two-photon state the kA$_g$ \cite{Shukla03}. Table 
I shows that, (i) exactly as near the 1B$_u$, a pair of excimers and a 
polaron-pair occur near the mA$_g$, and (ii) longitudinal
transition dipole-couplings between interchain states near the 1B$_u$ and 
interchain states of the {\it same character} near the mA$_g$, are of the {\it same
strength as that between the 1B$_u$ and mA$_g$ excitons.}

\begin{figure}
 \centering
 \includegraphics[clip,width=2.4in]{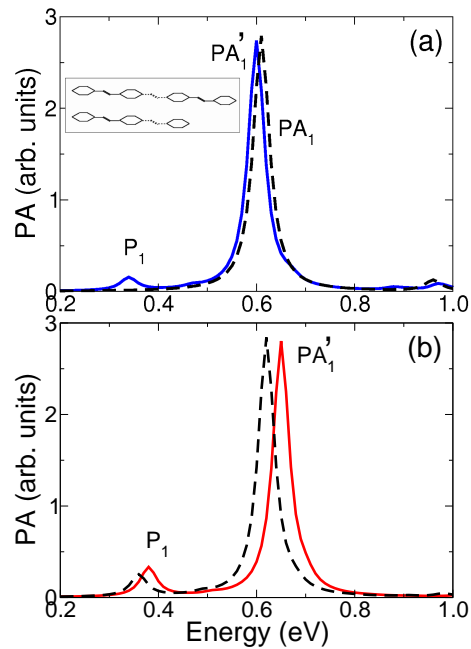}
 \caption{(a) Calculated PAs for a  
8-unit PPV oligomer (dashed curve), and for a two-chain system consisting of a
7-unit and a 9-unit oligomer, for $\kappa_{\perp}=2$ and $t_{\perp}=0.1$
eV. A common linewidth of 0.02 eV is assumed. 
The inset shows schematically the arrangement of
the oligomers, 
with the ends matching on one side only. (b)
Calculated PAs of the same two-chain system of (a), for $\kappa_{\perp}= 1.75$ and $t_{\perp}=0.15$
eV (dashed curve), $\kappa_{\perp}= 1.5$ and $t_{\perp}=0.2$
eV (solid curve). 
PA$_1$ (PA$_1^{\prime}$) is from the exciton (excimer).
}
\end{figure}

We now relax the inversion symmetry condition to take 
disorder into account approximately.
We consider two oligomers of different lengths, 
arranged face to face with only one end matching (see insert Fig.~2).
In Table II we show the
results of SCI calculations for PPV oligomers 5- and 7-units long, again for
interchain separation 0.4 nm. 
The key differences from the centrosymmetric case are: (i) there are
now two optical excitons ($j=3,4$), which have acquired weak
ionic character, (ii) characterization of states as excimers ($j=$ 2, 6, and 7)
and polaron pairs ($j=5$) are still valid, and
(iii) the excimers have acquired weak transition dipole couplings with the ground state.
We have shown the energy region near
the optical exciton only. Instead of giving similar details for higher energy
states, we will calculate PA.

The key results that emerge from Tables I and II are: (i) direct photogenerations of the
optical exciton and the two lowest excimers, one below and one above the exciton,
are allowed, and (ii) the excimers, and not free polarons
or the polaron-pair play a crucial role in PCP films. 
While we have shown calculations of interchain interaction effects for 
ideal and simple cases, 
our results remain
qualitatively intact for three or more oligomers, different relative orientations and 
distances. 

Our calculations therefore provide the foundation for understanding 
PCP films qualitatively, and perhaps even semiquantitatively.
The reduced PL and its nonexponential behavior in films are related to the
lowest excimer. Nonradiative relaxation of the exciton
to the lowest excimer competes with direct radiative relaxation. It is likely that
PL and PA in films after a few hundred fs are largely from the weakly emissive lowest
excimer. The allowed component in the excimer's wavefunction comes from the
optical exciton, making it likely that their emission profiles are similar.
This conjecture is supported by very recent observations of (i) 
weak emission from an intermolecular species in 
dendritic oligothiophenes \cite{Zhang07}, and (ii) PA from excimers in pentacene films
\cite{Marciniak07}.
Delayed PL
\cite{Rothberg06,Arkhipov04} is from both the excimers, 
with the upper excimer relaxing directly to the ground state 
as well as via the exciton.
The polaron-pair 
component of the excimers' wavefunctions 
ensures that the relaxation processes are slow, as the electron and the hole are partly
on different chains. The quenching of the delayed PL by external electric field  
is understandable. The 
transverse dipole
coupling between the two excimers and the polaron-pair indicates that an external electric
field will mix these energetically proximate states strongly \cite{theory-nlo},
as we have confirmed numerically (not shown). Increased ionicity
of the excimers' wavefunction diminishes 
both the radiative decay to the ground state and the nonradiative decay to
the exciton. This mechanism provides
a natural explanation for the immediate reappearance of the delayed PL upon removal of the field.
 
A direct test of our theory comes from comparisons of calculated and experimental PAs.
In Fig.~2(a) we compare PAs calculated for a single
8-unit PPV oligomer with that from the lowest excimer in a two-chain system
consisting of a 7-unit and a 9-unit oligomer. 
Fig.~2(b) shows PA calculations for the same two-chain system for other
parameters.
The similarity between the two-chain PA in Fig.~2 and the
experimental PA spectra of Sheng {\it et al.} \cite{Sheng07} for films is striking.
PA$_1$ in the single chain corresponds to the transition from the 1B$_u$ to the
mA$_g$. The initial and final states of PA$_1^{\prime}$ absorptions in the two-chain
systems are both excimers.
The near identical energies of this absorption
in solutions and films \cite{Sheng07}, and the 
absence of correlation between PA and PL decays in the latter \cite{Rothberg06,Yan94},  
can therefore both be explained. 

The P$_1$ absorption in the two-chain system in Fig.~2
is from the lowest excimer to the lowest
polaron-pair. Its strength is indeed a measure of interchain interaction \cite{Sheng07}, 
but it is unrelated to free polarons.
Our interpretation of P$_1$ resolves the apparent disagreement between 
ultrafast spectroscopy \cite{Sheng07} 
and other measurements \cite{Dicker04,Hendry05}. We predict
that the polarizations of P$_1$ and PA$_1^{\prime}$ are
different.

In summary, excimers and polaron-pairs occur not only near the
optical edge of PCP films, but also at high energies. The 1B$_u$
and mA$_g$
excitons, together with the excimers and polaron-pairs derived from these states 
constitute the essential
optical states of PCP films. We have
presented the first applications of this concept here. Application to copolymers 
\cite{Sreearunothai06} is clearly of interest.

S. M. thanks Z. V. Vardeny for suggesting this work and for the hospitality extended by
the University of Utah where this work was conceived. We are grateful to L. J. Rothberg
and C.-X. Sheng for many stimulating discussions and for sending preprints.
This work was supported by NSF-DMR-0705163.

\end{document}